\documentclass[apjl]{emulateapj}

\usepackage{graphicx}
\usepackage{natbib}
\usepackage{amsmath}
\usepackage{amssymb}
\usepackage{epstopdf}
\usepackage{xcolor}
\begin{document}

\title{An upper limit on the mass of the circum-planetary disk for DH~Tau~b\footnotemark[*]}\footnotetext[*]{This work is based on observations carried out under project D15AC with the IRAM NOEMA Interferometer. IRAM is supported by INSU/CNRS (France), MPG (Germany) and IGN (Spain).}

\author{Schuyler G. Wolff\altaffilmark{1}}
\email{swolff9@jh.edu}
\author{Fran\c cois M\'{e}nard\altaffilmark{2}}
\author{Claudio Caceres\altaffilmark{3}}
\author{Charlene Lef{\`e}vre\altaffilmark{4}}
\author{Mickael Bonnefoy\altaffilmark{2}}
\author{H{\'e}ctor C{\'a}novas\altaffilmark{5}}
\author{S\'ebastien Maret\altaffilmark{2}}
\author{Christophe Pinte\altaffilmark{2}}
\author{Matthias R. Schreiber\altaffilmark{6}}
\author{Gerrit van der Plas\altaffilmark{2}}

\altaffiltext{1}{Department of Physics and Astronomy, Johns Hopkins University, Baltimore, MD 21218, USA}
\altaffiltext{2}{Universit\'e Grenoble-Alpes, CNRS, Institut de Plan\'etologie et d'Astrophyisque (IPAG), F-38000 Grenoble, France}
\altaffiltext{3}{Departamento de Ciencias Fisicas, Facultad de Ciencias Exactas, Universidad Andres Bello. Av. Fernandez Concha 700, Las Condes, Santiago, Chile.}
\altaffiltext{4}{IRAM, 300 rue de la piscine, F-38406 Saint-Martin-d’H\`eres, France}
\altaffiltext{5}{Departamento de F\'isica Te\'orica, Universidad Aut\'onoma de Madrid, Cantoblanco, 28049 Madrid, Spain.}
\altaffiltext{6}{Instituto de F\'isica y Astronom\'ia, Universidad de Valpara\'iso, Blanco 951, Valparaíso, Chile}




\begin{abstract}
DH~Tau is a young ($\sim$1 Myr) classical T~Tauri star. It is one of the few young PMS stars known to be associated with a planetary mass companion, DH~Tau~b, orbiting at large separation and detected by direct imaging. DH~Tau~b is thought to be accreting based on copious H${\alpha}$ emission and exhibits variable Paschen Beta emission. NOEMA observations at 230 GHz allow us to place constraints on the disk dust mass for both DH~Tau~b and the primary in a regime where the disks will appear optically thin. We estimate a disk dust mass for the primary, DH~Tau~A of $17.2\pm1.7\,M_{\oplus}$, which gives a disk-to-star mass ratio of 0.014 (assuming the usual Gas-to-Dust mass ratio of 100 in the disk). We find a conservative disk dust mass upper limit of 0.42$M_{\oplus}$ for DH~Tau~b, assuming that the disk temperature is dominated by irradiation from DH Tau b itself. Given the environment of the circumplanetary disk, variable illumination from the primary or the equilibrium temperature of the surrounding cloud would lead to even lower disk mass estimates. A MCFOST radiative transfer model including heating of the circumplanetary disk by DH~Tau~b and DH~Tau~A suggests that a mass averaged disk temperature of 22 K is more realistic, resulting in a dust disk mass upper limit of 0.09$M_{\oplus}$ for DH~Tau~b. We place DH~Tau~b in context with similar objects and discuss the consequences for planet formation models.
\end{abstract}

\section{Introduction}
\label{Sec:intro}

With well over 3000 confirmed extrasolar planets now known, the focus of exoplanet science is shifting from their discovery to understanding the details of their formation and evolution. However, increasing our understanding of this complex process can only be achieved with unambiguous detections of planetary mass bodies still in formation. Today, a handful of good candidates are known \citep{2012ApJ...745....5K,2015Natur.527..342S,2014ApJ...792L..22B,2014ApJ...792L..23R,2015ApJ...807...64Q}, but they are still embedded deeply in the circumstellar disk and also located close to the central objects. These are challenging conditions to study the processes that lead to their formation.

Fortunately, a small population of planetary mass companions (PMCs) has recently been discovered that offers a much better opportunity to study the planet formation process in greater details with current instruments. These PMCs, identified by direct imaging surveys in the NIR, orbit very young host stars (T Tauri stars) and they do so at large enough separations to be easily observable, typically several hundred au ($\sim$ 1 arcsec) \citep[e.g., ][]{2005A&A...435L..13N,2008ApJ...689L.153L,2008A&A...491..311S,2011ApJ...726..113I,2014ApJ...780L...4B}.

While planets at separations of $<$ 100 au are thought to be the consequence of either core accretion \citep{2007prpl.conf..591L} or gravitational instabilities \citep{1997Sci...276.1836B,2011ApJ...731...74B} acting at the Class II stage (i.e., T Tauri stage), planets at larger separations are believed to be products of disk fragmentation at an earlier stage \citep[Class 0 or I stage, ][]{2010ApJ...710.1375K}. All these mechanisms require that a forming planet builds up from its own circumplanetary disk that formed either from the surrounding cloud, or from the massive disk around the host star.
Indirect evidence for the presence of such disks is provided by the fact that planet-mass companions in young systems are powerful H${\alpha}$ emitters, e.g., OTS 44, GSC 06214-00210 b, GQ Lupi b, FW Tau c, DH Tau b \citep{2013A&A...558L...7J,2014ApJ...783L..17Z}. The H${\alpha}$ emission, or some portion of it, being the trace of accretion from the disk onto the companion.
The more direct detectability of these circumplanetary disks was recently demonstrated when ALMA measured the continuum and CO emission around the PMC orbiting the TTauri binary FW Tau \citep{2015ApJ...806L..22C}. The disk around FW Tau C (the PMC) has an estimated disk mass of 2-3 $M_{\oplus}$. 
Attempts have been made to resolve the circumplanetary disks around several other PMCs with radio interferometer (e.g., GSC 0614-210 B; Bowler et al.~2015, GQ Lupi; Dai et al.~2010, MacGregor et al. 2016), but no other detections exist to date.

\begin{figure}[!thb]
\begin{center}
\includegraphics[width=\columnwidth]{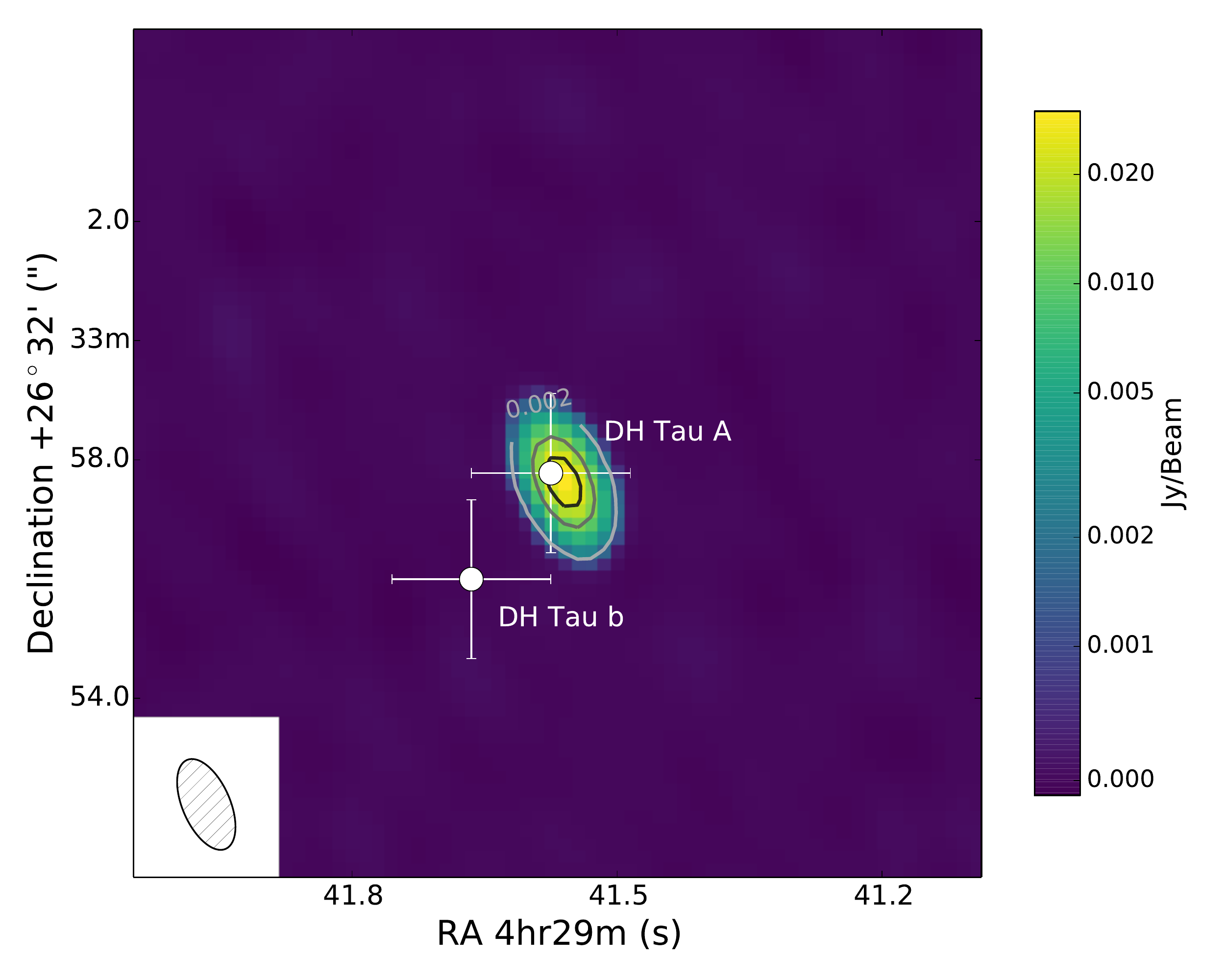}
\caption{1.3 mm continuum NOEMA observations of the DH Tau system. The disk of DH Tau A is clearly detected, but is unresolved. The disk of DH Tau b is undetected. Contours are drawn beginning at 0.002 Jy/beam in intervals of 0.01 Jy/beam. The errors in the positions of the DH Tau A and b components are dominated by the proper motion uncertainties. The symmetric sidelobes are an artifact of the baseline configuration. The inset shows the beam with a PA of 27.8$^{\circ}$, a major axis of 1.61,'' and a minor axis of 0.79.''
\label{fig:selfcal}}
\end{center}
\end{figure}

\begin{center}
The DH Tau System
\end{center}
DH Tau is a binary system with a projected separation of 330 AU (2.\!''3). The system is located in the Taurus star forming region at a distance of 140 pc, with an extinction in the \textit{J} band of $A_{J} = 0.3 \pm 0.3$ \citep{2005ApJ...620..984I}, and a mean age of 2.3 Myrs \citep{2007A&A...473L..21B}. The primary is a classical T Tauri star with an M1Ve spectral type \citep{1977ApJ...214..747H} with $\log(T/K) = 3.5688 \pm 0.0170$ and $\log(L/L_{\odot}) = -0.262 \pm 0.110$  \citep{2013ApJ...771..129A}. DH Tau b was initially discovered by \citet{2005ApJ...620..984I}, who classified it as a L2 spectral type brown dwarf with a mass of $30 - 50 \, M_{Jup}$. \citet{2006ApJ...649..894L} later compared bolometric luminosities to updated evolutionary tracks and gave a revised mass estimate of $11^{+10}_{-3} \, M_{Jup}$, placing it near the exoplanet/brown dwarf boundary. \citet{2012A&A...540A..85P} modeled the atmosphere using \textit{J, H,} and \textit{K} spectra, and inferred a radius for DH Tau b of $2.7 \pm 0.8 \, R_{Jup}$, and a temperature of $2350 \pm 150$ K. \citet{2014A&A...562A.127B} give a spectral type for DH Tau b of M9.25$\pm$0.25 (corresponding to $15 \, M_{Jup}$).

DH Tau b is the youngest PMC known to date.
It is known to be actively accreting, as traced by very strong H$\alpha$ emission \citep{2014ApJ...783L..17Z}. 
The Pa${\beta}$ line of Hydrogen is also reported, in emission, by \citet{2014A&A...562A.127B} further supporting the idea that DH Tau b is still accreting.
DH Tau as a system also displays unresolved MIR excess which, given the accreting nature of DH Tau b, is likely caused in part by the circumplanetary disk. 
\citet{2012ApJ...751..115H} reported a 47 mJy detection around the DH Tau primary at 0.88 mm.  Their observations with the SMA only provided a 3$\sigma$ upper limit of 10 mJy at 0.89 mm for DH Tau b. The circumplanetary disk has remained undetected to date. 

In Section \ref{Sec:obs}, we present the NOEMA observations of the DH Tau system and the VLT/SINFONI spectroscopy of the Pa${\beta}$ Hydrogen line. Section \ref{Sec:model} presents the upper limits on the disk mass of DH Tau b, an estimated disk mass for DH Tau A, and the disk model used. Finally, in Section \ref{Sec:dis} we discuss the disk mass results and place them in context with other observations of circumplanetary disks.

\section{Observations}
\label{Sec:obs}

\subsection{NOEMA 1.3mm continuum imaging}
\label{Sec:obsnoema}

The data presented in Fig.~\ref{fig:selfcal} were obtained with NOEMA, the NOrthern Extended Millimeter Array. The observations were carried out on December 10th, 2015. At that time the array was in the 7C compact configuration, with 6 antennas operating. Station W09 was off-line. Antennas were based on stations E12, N17, N11, E18, W12, and E04. The resulting 15 baselines ranged from 48m to 240m in length (unprojected). DH Tau and its companion were observed for a total of 6.5 hours between hour angle -0.3h and +6.0h, of which 4.5 hours we spent on-source. The rest of the time was used for calibration.   

We used 0400+258 and 0507+179 as phase calibrators. The atmospheric conditions were excellent and the rms phase noise was measured between 12$^o$ on short baselines and 29$^o$ on long baselines, at 1.3 mm. This phase noise introduces a position error of less than 0.1 arcsec. The source LkHa 101 was used for the flux calibration, while 3C84 was used for the bandpass calibration. We consider an absolute flux uncertainty of 10\%. The total bandpass for the 230.5 GHz continuum measurement was 3.2 GHz in each polarisation. We excluded a short range (80 MHz) that included the CO(2-1) line. The GILDAS software package was used to reduce the data. 

The continuum map was produced using natural weighting of the visibilities to favour signal-to-noise over angular resolution. The resulting beam size is $1.\!"61 \times 0.\!"79$ at P.A. 28$^o$. High signal-to-noise on DH Tau A allows for phase self-calibration. This allowed us to correctly remove side-lobes that remained present after the first reduction steps. We do not perform amplitude self-calibration in order to preserve the absolute flux measurement. The phase self-calibration stopping criteria was set to 1000 iterations. From the visibilities, we compute the stable thermal noise limit (absolute flux limit) to be 0.0653 mJy. Any residuals after self-calibration correspond to a lack of uv coverage that is impossible to correct.


We find a $1 \sigma$ flux limit of 0.0653 mJy/beam for the 1.3 mm continuum data, which corresponds to a $3 \sigma$ upper limit for the DH Tau b circumplanetary disk flux of 0.196 mJy. Primary beam attenuation was not taken into account because of the small separation between DH Tau A and DH Tau b (beam attenuation $<$2\% at the position of DH Tau b). We detect the central component of the system, DH Tau A, at $>100\sigma$, with an integrated disk flux of $30.8 \pm 0.2$ mJy. 


\begin{figure}[b]
\begin{center}
\includegraphics[width=3.0in]{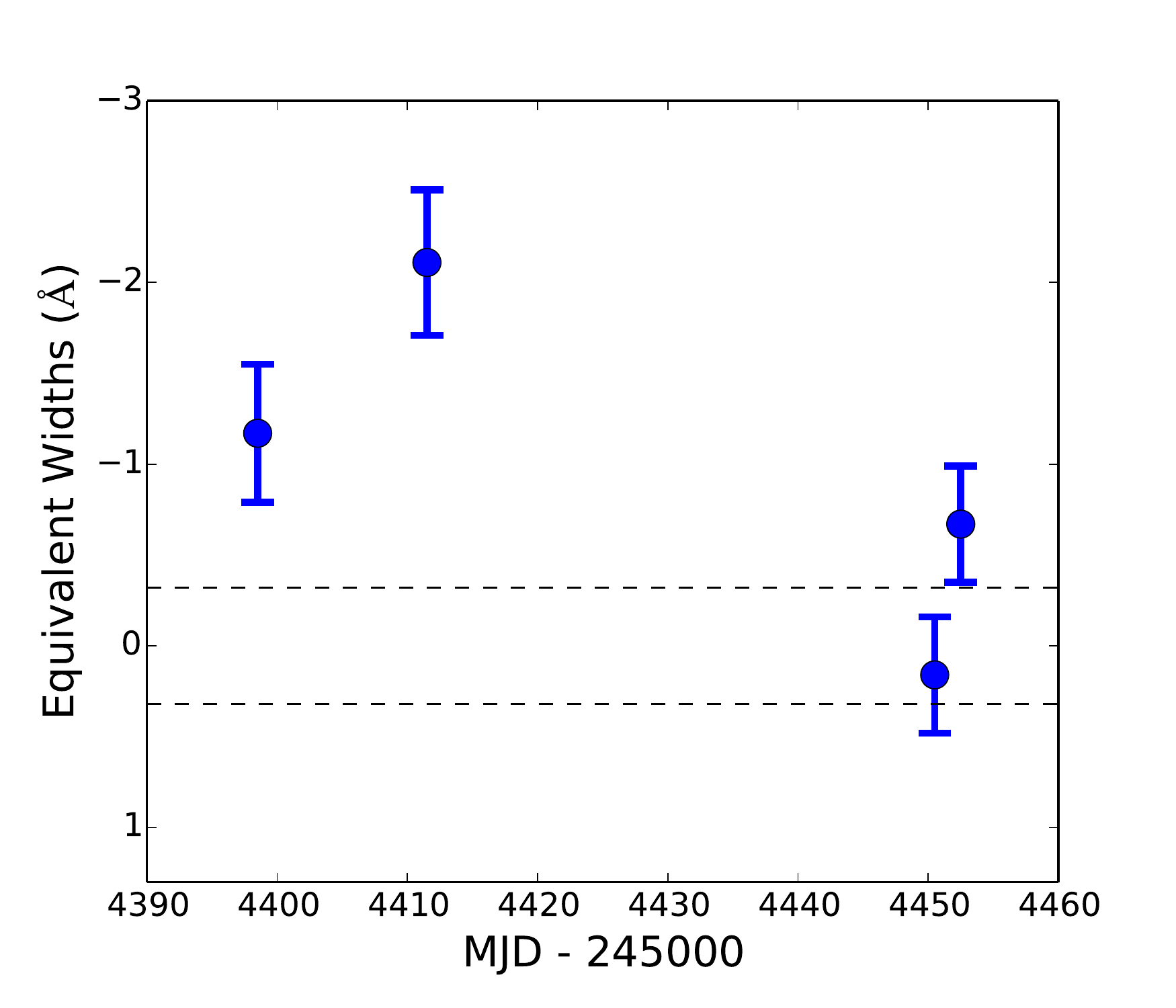}
\caption{Variability of the Pa $\beta$ equivalent width with time provides further evidence of an accreting circumplanetary disk surrounding DH Tau b. The dashed lines represents the mean $1\sigma$ error for the equivalent width measurements. \label{fig:EWevolution}}
\end{center}
\end{figure}

\subsection{VLT/SINFONI spectroscopy of the Paschen $\beta$ Hydrogen line}
\label{Sec:obssinfoni}


DH Tau b was observed with the VLT/SINFONI instrument on Oct. 25th, Nov. 7th, Dec. 16th, and Dec. 18th, 2007 (program ID 080.C-0590(A)). SINFONI is composed of an integral field spectrograph SPIFFI fed by the adaptive optics module MACAO \citep{2003SPIE.4841.1548E, 2004SPIE.5490..130B}.  The instrument was operated with the J-band grating yielding a spectral resolution of $\sim$2000 over the 1.1-1.35 $\mu$m range. The pre-optics was sampling the 0.8$\times$0.8'' field-of-view with a spaxel size on sky of 12.5$\times$25 mas. Each sequence is composed of 8$\times$300s exposures with small dithering and one acquisition on the sky at the end to ensure a proper removal of the sky emission. Telluric standard stars were observed after DH Tau on each night to estimate the contamination by telluric features in the companion spectra. Because the Paschen $\beta$ line is not significantly affected by telluric lines in our spectra, we decided not to correct for telluric features in order to avoid adding noise to our spectra. The October, November, and December 16  data were published in \cite{2014A&A...562A.127B}. We reduced the December 18 data with the same tools as used in \cite{2014A&A...562A.127B} in order to get a homogeneous set of extracted spectra of the companion.  

A Paschen $\beta$ emission line is detected in the November observations, marginally detected in October, and not detected in December. All spectra have a comparable estimated S/N between 1.29 and 1.31 $\mu$m\footnote{The S/N was computed in a two step process. We first interpolated the IRTF spectrum of the M9 dwarf LP 944-20 on the SINFONI wavelength grid and normalized it in flux to the flux of the pseudo-continuum of DH Tau b over the 1.29-1.31$\mu$m range. We used this template spectrum to approximate, then remove, all the intrinsic features of the DH Tau b (FeH lines mostly) and compute the local level of the noise.}. We estimated the equivalent width of the line following the method of \cite{1992ApJS...83..147S}. The continuum was estimated in a range adjacent to the line, between 1.277 and  1.281 $\mu$m, and between 1.283 and 1.287 $\mu$m. The equivalent width is computed between 1.281 and 1.283 $\mu$m. The values are reported in Table \ref{Tab:EWPaB} and their evolution in time is shown in Figure \ref{fig:EWevolution}. 

Assuming the Pa${\beta}$ line in emission is tracing accretion of material onto DH Tau b, then the results presented in
Table \ref{Tab:EWPaB} and Figure \ref{fig:EWevolution} provide indications that the accretion process itself may be variable in time.
This is reminiscent of the well documented variability of the accretion process in more massive T Tauri stars, e.g., \cite{2016A&A...586A..47S}. The poor time coverage for the spectral variations of DH Tau b forbids a deeper analysis. We do not discuss further the variability of accretion in DH Tau b, but note it is very likely present.

\begin{table}
\caption{Equivalent width of the Paschen $\beta$ line}             
\label{Tab:EWPaB}      
\centering                          
\begin{tabular}{ cccc}        
\hline \hline                 
MJD Date - 245000 & S/N & Eq. Width ($\AA$) \\
\hline                        
4398.5	 &	16	 &	-1.17$\pm$0.38 \\
4411.5	&	18	&	-2.11$\pm$0.40 \\
4450.5	&	20	&	0.16$\pm$0.24 \\
4452.5	&	19	&	-0.67$\pm$0.32 \\
\hline                 
\end{tabular}
\end{table}

\section{Circumplanetary Disk Models and Results}
\label{Sec:model}
 
In this section we present models for the dust mass estimates extracted from the 1.3 mm continuum NOEMA data. 
We will consider three cases: the disk of DH Tau b is heated by DH Tau b only; the disk is heated by DH Tau A; and the disk is in equilibrium with the ambient cloud (assumed at 20 K). To test the dominant source of the disk dust temperature, we combine the contributions from DH Tau A and DH Tau b using a radiative transfer model.

We expect the disk to be optically thin at 1.3 mm. In this case, the disk dust mass can be expressed as
  \begin{equation*}
     M_{dust} = \frac{F_{\nu} D^{2}}{\kappa_{\nu} B_{\nu}(T_{\mathrm{disk}})}  
 \end{equation*}
where $F_{\nu}$ is our measured $\nu = 230 \, GHz$ (1.3 mm) 3$\sigma$ flux limit, D is the distance (140 pc), $\kappa_{\nu}$ is the dust opacity, and $B_{\nu}(T_{\mathrm{dust}})$ is the Planck function evaluated at the disk temperature. We use the dust opacity law from \citet{1990AJ.....99..924B}; $\kappa_{\nu} = 10 \, (\nu / 10^{12} Hz)^{\beta} \, cm^{2} g^{-1} = 2.3 \, cm^{2} g^{-1}$ for frequency, $\nu$, and power law index, $\beta = 1$.

Typical disk temperatures are $\sim 20$ K, but this varies with stellar luminosity. For DH Tau b, we calculate a luminosity of 0.0021 $L_{\odot}$ using the radius and stellar temperature \citep{2014ApJ...783L..17Z}. \citet{2016ApJ...819..102V} provide a scaling relation between stellar luminosity and disk temperature for low mass stars; $T_{\mathrm{disk}} = 22 (L/L_{\odot})^{0.16} \, K$, which gives a disk temperature for DH Tau b of 8.2 K. 

It is worth noting here that the temperature of molecular clouds is typically in the range of 10 - 20 K \citep{1987ASSL..134...51G}. In this case, the temperature of the disk may depend more on the ambient temperature from the Taurus SFR than the central source. Likewise, DH Tau b is located nearby to the much more luminous DH~Tau~A primary, with a luminosity of $0.55 \, L_{\odot}$ and an effective stellar temperature of $T_{*} = 3706$ K \citep{2013ApJ...771..129A}. If we treat the dust as a blackbody in thermal equilibrium with the central star, DH Tau A, at the distance of DH Tau b (330 au) with a circumplanetary disk albedo of a = 0.5, we expect the equilibrium temperature to be $T = T_{*}(1-a)^{1/4} \sqrt{R_{*}/2D} = 11 \, K$. Depending on the orientation of the disk relative to the central star and/or the optical depth of the disk, there could be some additional heating due to illumination from the primary, DH~Tau~A. 
Viscous heating due to accretion could also raise the temperature of the disk and serve as another source of uncertainty in the dust mass estimate.

 \begin{figure}[!thb]
\begin{center}
\includegraphics[width=0.5\textwidth]{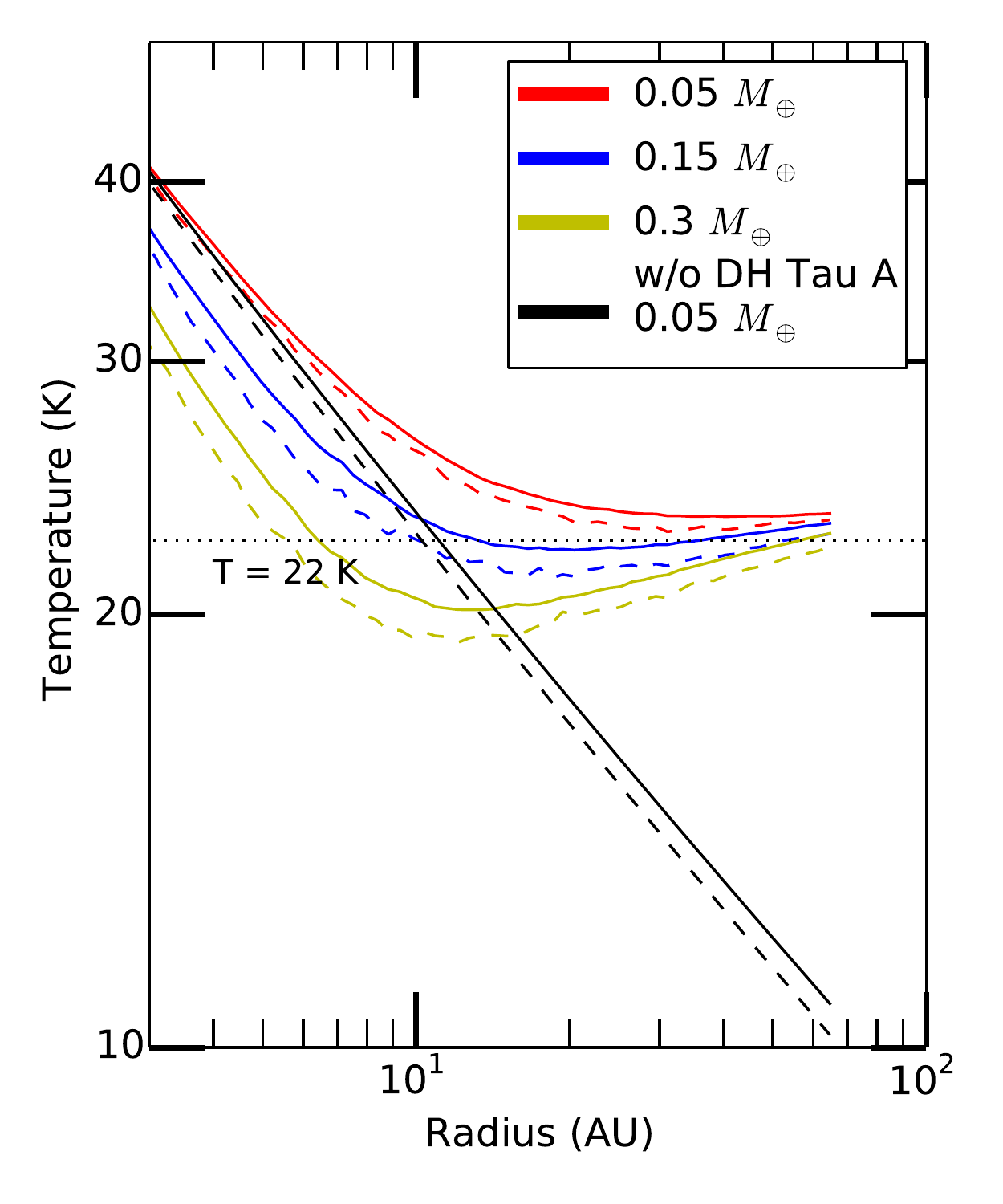}
\caption{Dust temperature profile for the set of MCFOST radiative transfer disk models with different disk dust masses. The dashed lines show the radial profile of the disk midplane temperature, while the solid lines show the radial profile of the mass-averaged dust temperature. As the mass increases, the disk becomes more optically thick to radiation and the temperature decreases. At the outer edges of the disk, all dust mass models converge to a mass-averaged dust disk temperature of 22 K (as indicated by the dotted line).  \label{fig:mcfost}}
\end{center}
\end{figure}

 We reproduce the effect of the host star on the disk dust temperature by generating an MCFOST radiative transfer model of the system \citep{2006A&A...459..797P,2009A&A...498..967P}. MCFOST is a Monte Carlo Radiative Transfer code designed to study circumstellar disks. At each grid location in the modeled disk, the temperature and scattering source function are computed via a Monte Carlo method: photon packets are propagated stochastically through the model volume following the equations of radiative transfer. MCFOST allows the user to include multiple radiative sources, allowing the inclusion of the DH Tau A primary located 330 au from the circumplanetary disk. DH Tau A was modeled using an effective temperature of 3700 K and a low surface gravity of $\log(g) = 3.5$, while DH Tau b was assumed to have an effective temperature of 2300 K with $\log(g) = 3.5$. For DH Tau b, we assume an axisymmetric disk model with a gas supported flaring exponent of 1.125, and a surface density described by a power law in radius with an index of -0.5. The grains are comprised of astronomical silicates with a grain size distribution defined by an ISM-like -3.5 power law exponent and grain sizes ranging from 0.1 to 1000 $\mu m$. The resulting dust opacity is 2.29 $cm^{2}/g$, similar to the dust opacity of 2.3 $cm^{2}/g$ predicted above for an optically thin disk. We assume a gas-to-dust mass ratio of 100 and a distance of 140 pc. For simplicity DH Tau b receives light directly from DH Tau A without attenuation, as if DH Tau b was located out of plane from the disk of DH Tau A. 
 
 The typical value for the outer radius of a circumplanetary disk is not well constrained. Numerical simulations of embedded circumplanetary disks suggest that the radii truncate at a fraction of the Hill radius due to interactions with the viscous, young circumstellar disk \citep{2010AJ....140.1168W}. For DH Tau b, the Hill radius is $R_{\mathrm{Hill}} = a (\frac{M_{p}}{3 M_{*}})^{1/3} \simeq 70$ au for the planetary mass and separation ($M_{p} = 11 \, M_{Jup}$, $a = 330$ au respectively) given in the introduction, and the primary star mass ($M_{*}$) of $0.37 \pm 0.12 \, M_{\odot}$ from \citet{2005ApJ...620..984I}. 
 Alternatively, if the DH Tau b disk formed from the collapse of the surrounding cloud, it would be expected to have a larger radius. \citet{2009ApJ...701..698S} survey the disks of young, low mass stars in the Taurus-Auriga star-forming region and find a range in disk outer radii of $R_{\mathrm{out}} \sim 100 - 1000$ au from resolved CO emission.
 However, a larger disk would be truncated due to the presence of the primary at a $\sim$ 0.3 - 0.5 fraction of the 330 au separation \citep{1977MNRAS.181..441P}. A disk truncated at 110 au (diameter $\sim 1.6''$) is roughly the same size as the beam along the major axis ($1.61''$). 
 In either case, this is below the beam size, and we treat the disk as a point source in our data. For the MCFOST model, we define the disk outer radius to be the Hill radius of 70 au. 
 
The model was tested for several disk dust masses covering the range predicted for the various dust temperatures, and for different orientations of the circumplanetary disk with respect to DH Tau A. We found that the disk orientation (e.g., face-on, edge-on, or intermediate illumination from the central star) has no measurable effect on the azimuthally averaged dust temperature.  Figure \ref{fig:mcfost} shows the mass-averaged temperature profile for three assumed disk masses and includes for comparison a disk model without the host star included. As the disk dust mass increases, the mid-plane temperature of the disk decreases as the disk becomes more opaque to radiation. In the outer regions of the disk, all MCFOST models converge on a dust disk mass-averaged temperature of $22 \pm 2 \, K$, corresponding to a disk dust mass upper limit of $0.09 \pm 0.01 \, M_{\oplus}$. This temperature and associated mass estimate is more consistent with what is expected for a disk in a young star forming region.  We caution that assumptions in the disk model and the uncertainty in the separation of DH Tau b could result in a lower dust temperature.

Table \ref{table:mass} gives the estimated disk mass upper limits for DH Tau b for the various temperatures described above. On the most conservative end, we provide an upper limit on the circumplanetary disk mass of 0.42 $M_{\oplus}$. However, the dust disk mass can likely be constrained further given the circumplanetary environment and as suggested by radiative transfer models of the system to be 0.09 $M_{\oplus}$. We adopt this upper limit for future discussion.

The temperature derived disk masses quoted above assume that the disk is optically thin at the 1.3 mm wavelength. If the disk were optically thick, i.e., $\tau > 1$ where $\tau = \int \rho \kappa_{\nu} ds = \kappa_{\nu} \Sigma > 1$, then the observed flux can be used to set a lower limit on the extent of the disk. Using the dust opacity law given above with $\beta \simeq 0$ for the optically thick case, the DH Tau b disk dust mass of 0.09 $M_{\oplus}$, and assuming a flat surface density we can constrain the radius of the disk: $R < \sqrt{\kappa_{\nu} \, M_{D} / \pi} < 2.9$ au. Therefore, if the disk were optically thick, it would have to be compact.

Using the same formalism with a midplane disk temperature of 20 K as predicted from the van der Plas et al. (2016) stellar luminosity relation for low mass stars, we estimate a disk dust mass for the primary, DH Tau A of $17.2 \pm 1.7 \, M_{\oplus}$. The uncertainties are based on the absolute flux uncertainty and do not include errors in the assumed distance and disk opacity.

\begin{table}
\centering
\caption{DH Tau b Disk Dust Mass Upper Limits}
\begin{tabular}{lll}
\hline\hline
Temp. & Dust Mass Limit & Source \\
\hline
20 K & $0.11 \pm 0.01 \, M_{\oplus}$ & Ambient cloud Temp. \\ 
8.2 K  & $0.42 \pm 0.04 \,  M_{\oplus}$ & DH Tau b Luminosity \\
11 K & $0.26 \pm 0.03 \,  M_{\oplus}$  & Illumination from primary \\
\textbf{22 K} & \boldmath{$0.09 \pm 0.01 \, M_{\oplus}$ } & \textbf{MCFOST model}\\
\hline
\label{table:mass}
\end{tabular}
\end{table}

\section{Discussion}
\label{Sec:dis}

We are able to place an upper limit on the circumplanetary disk mass of the DH Tau b PMC. While the dust mass limit of $0.09 \, M_{\oplus}$ is clearly not massive enough anymore to form planets, it still provides $\sim 8$ lunar masses of solid material to form satellites or minor bodies orbiting DH Tau b. The circumstellar disk surrounding DH Tau A has a dust mass of $17 \, M_{\oplus}$, which is above the limit required to form giant planet cores ($\sim 10 \, M_{\oplus}$), and could still support the formation of several terrestrial planets. The circumstellar disk mass is comparable to other Taurus disk masses for this spectral type, with a disk to star mass ratio of 0.014, assuming a gas to dust ratio of 100. The equivalent disk to star mass ratio for DH Tau b would require a total disk mass of $\sim48 \, M_{\oplus}$, which is not reproduced by even our most conservative detection limit for an uncharacteristically low mass averaged dust temperature. 

For DH Tau b, the mass accretion rate predicted from H$\alpha$ observations is $3.2 \times 10^{-12} \, M_{\odot}/yr$ \citep{2014ApJ...783L..17Z}. Using the disk mass limit derived from the MCFOST model gives a disk dissipation timescale of 9.1 Myrs assuming a gas-to-dust ratio of 100. 



\subsection{Comparison to Known PMC Disk Masses}

While there are not yet many disk mass estimates using millimeter continuum data for planetary mass objects, we compare these estimations for DH Tau b with the results for three other known wide separation PMCs: FW Tau C, GSC 6214-210 B, and GQ Lup B. 

\begin{description}

\item[FW Tau] The FW Tau primary is actually a binary system with two M5 stars orbiting at 11 AU, while the companion, FW Tau C has a mass of 7 $M_{Jup}$ \citep{2015ApJ...798L..23K}. It is also in the Taurus SFR, with a similar age to DH Tau. Millimeter observations of the FW Tau system do not detect the circumbinary disk, but do detect the circumplanetary disk with an estimated dust mass of $\sim 2 \, M_{\oplus}$ \citep{2015ApJ...806L..22C}. This dust mass is well above the average dust to stellar mass ratio for the Taurus SFR. The non-detection of the primary disk is unusual, though it is possible that the binary system caused the circumbinary disk to dissipate more quickly. 

\item[GQ Lup] GQ Lup is in the Lupus 1 SFR, with a slightly older 3 Myr age \citep{2008A&A...489..143L, 2014A&A...561A...2A}. \citet{2010AJ....139..626D} conduct SMA 1.3 mm observations of GQ Lup (a young, 1 Myr old T Tauri star)  and detect the primary circumstellar disk with a mass of 3 $M_{Jup}$, but were unable to detect any disk signature around the secondary component. 
Recently published ALMA observations \citep{2016arXiv161106229M} detect a compact ($R_{\mathrm{out}} = 59 \pm 12$ au) circumprimary disk with a higher dust mass estimate of $\sim \, 15 \, M_{\oplus}$ from 870 $\mu m$ continuum observations. The circumplanetary disk is not detected with a $3 \sigma$ noise floor of 0.15 mJy/beam (equivalent to DH Tau b uncertainty) with a corresponding dust mass limit of $ < 0.004 \, M_{\oplus}$ calculated assuming that the dominant disk heating source is the primary. \citet{2016arXiv161106229M} also obtain $^{12}$CO and $^{13}$CO emission showing a gas disk that extends outside of GQ Lup b. A recent multi-wavelength study of the GQ Lup system using both ALMA continuum observations and MagAO optical photometry of the companion show that the circumstellar disk of GQ Lup A is misaligned with the spin axis, possibly due to interaction with GQ Lup b \citep{2017ApJ...836..223W}.

\item[GSC 0614-210] The circumplanetary disk around the 10 Myr old GSC 0614-210 B was not detected in ALMA continuum observations at 880 $\mu m$ with a 3$\sigma$ rms noise level of 0.22 mJy/beam \citep{2015ApJ...805L..17B}. This is comparable to the noise floor in these DH Tau observations, implying a similarly low disk mass ($<0.15 \, M_{\oplus}$). The circumstellar disk for the primary was not detected. 

\end{description}

\begin{figure}[t]
\begin{center}
\includegraphics[width=3.5in]{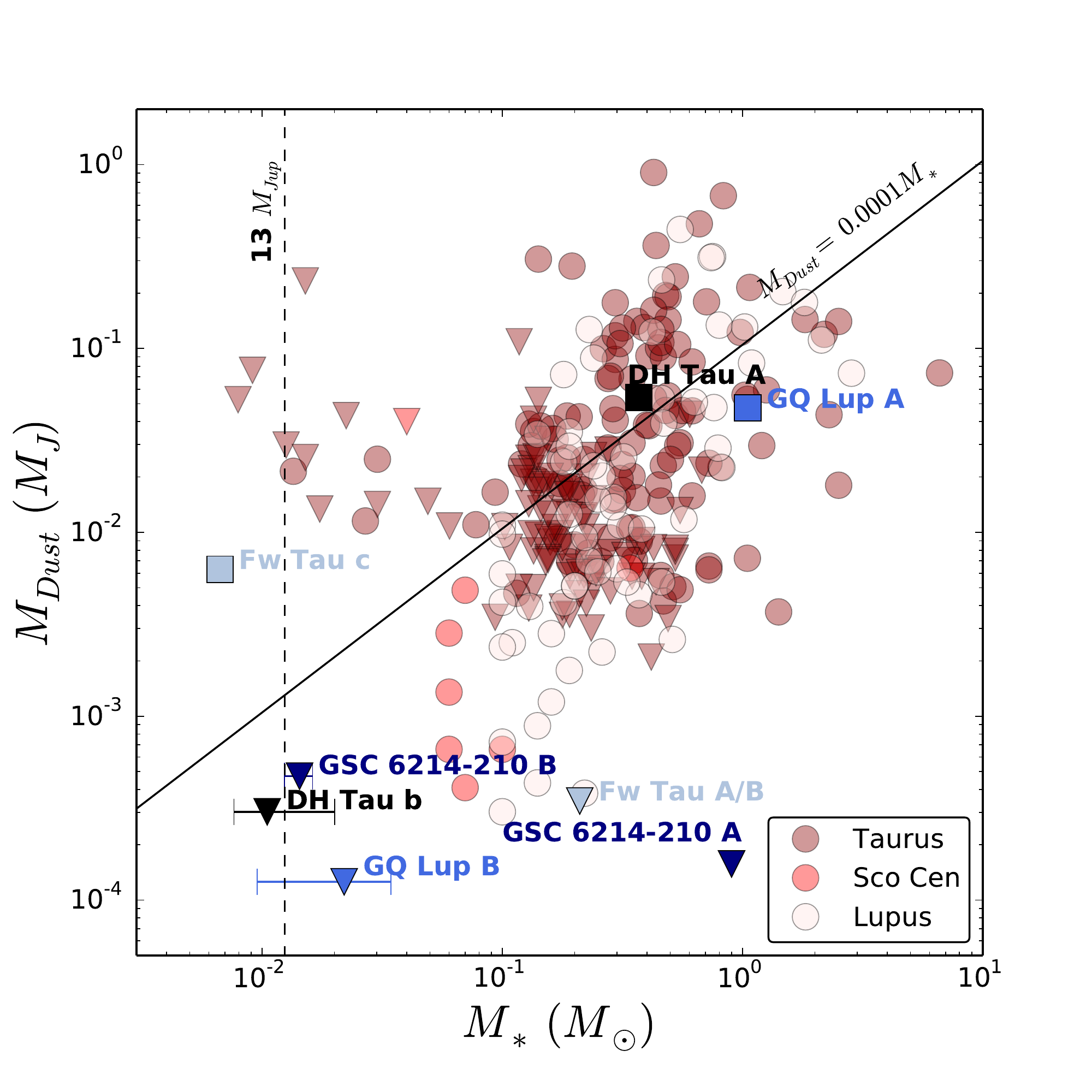}
\caption{The disk dust mass and stellar mass are shown for a collection of labeled PMCs with dust mass estimates from mm observations. Dust to star mass ratios are shown in red for a collection of stars in the Taurus \citep{2013ApJ...771..129A}, Lupus \citep{2016ApJ...828...46A}, and Sco Cen \citep{2016ApJ...819..102V} star forming regions. 3$\sigma$ upper limits are represented with triangles. The dashed vertical line represents the $13 \, M_{Jup}$ mass deuterium burning limit, while the solid diagonal line represents a 0.0001 $M_{*}/M_{Dust}$ ratio. The disk dust mass estimates for the PMCs are generally lower than expected for the mass of the object with the exception of FW Tau C which has an exceptionally large dust mass. \label{fig:dustmass_ratio}}
\end{center}
\end{figure}

The dust masses and stellar/planetary masses for the objects listed above are shown in Figure \ref{fig:dustmass_ratio}. In the event of a non-detection in the millimeter, 3$\sigma$ upper limits are provided. Included for comparison are the disk dust masses and stellar masses for a collection of objects in the Taurus \citep{2013ApJ...771..129A}, Lupus \citep{2016ApJ...828...46A}, and Sco Cen \citep{2016ApJ...819..102V} star forming regions. While the authors report the dust masses for the Lupus and Sco Cen circumstellar disks, the Taurus dust masses were computed from the provided mm fluxes using the stellar luminosity and temperature relation described in \citet{2013ApJ...771..129A}.

\subsection{Formation Mechanism?}

We use the ensemble of known PMCs to place constraints on the planet formation process. Different formation pathways should produce different signatures in both the accretion rates and the planet to dust disk mass ratios as compared to their environments.
Here we discuss the implications of the possible formation of these wide separation PMCs. 

\begin{itemize}

    \item[--] Disk Instability: 

    Models of giant planets produced via disk instabilities have difficulty producing massive planet cores outside of 100 au in all but the most massive disks. \citet{2013A&A...552A.129V} find that a protostellar disk mass of $\simeq 0.2 \, M_{\odot}$ is needed to produce planetary embryos with masses in the range of 3.5 - 43 $M_{Jup}$. DH Tau A and GQ Lup A have dust disk to star mass ratios below average, while the circumstellar disks for FW Tau A/B and GSC 6214-210 A were not detected at all. As the oldest system, it is possible that the GSC 6214-210 A disk has already dissipated in its 10 Myr lifetime. However, this formation scenario is difficult to support with the current low disk mass estimates. 

    \item[--] Core Accretion + Scattering:
    
    The most commonly employed formation mechanism for gas giant planets is via core accretion of pebbles in the parent protoplanetary disk. However, generating giant planet cores massive enough to accrete gas in situ at wide separations requires timescales longer than the lifetimes of the gas in the disk \citep{2007prpl.conf..591L}. 
    Alternatively, these planets could have been formed closer to their central stars and been dynamically scattered out to wider separations.
    None of the PMCs discussed here show evidence for a massive companion capable of dynamically scattering the PMC to wide separations. Indeed, direct imaging surveys of other wide separation PMCs do not find evidence for additional massive scattering companions and the core accretion + scattering event seems unlikely \citep[e.g.][]{2016ApJ...827..100B}. Surveys for scattering companions are limited by observational biases and this scenario cannot be ruled out. 
    
    A planet formed closer in but that has experienced such a dynamical scattering event that would now place it at a wide separation could potentially disrupt any circumplanetary disk. This scenario is supported by the low dust disk masses measured for all PMCs except FW Tau C, though their accretion signatures indicate that these disks are not entirely disrupted.
    Further monitoring of these systems to look for companions and/or signatures in their orbital properties indicative of a turbulent past could provide support to the core accretion + scattering model.

    \item[--] Turbulent Fragmentation of the molecular cloud:
    
    Through the process of turbulent fragmentation, filaments within dense molecular clouds gravitationally collapse to form protostellar/planetary cores as small as a few Jupiter masses \citep{1976MNRAS.176..367L}.
    While this formation mechanism is capable of forming low mass objects and has been invoked to explain the formation of free-floating brown dwarfs, it is difficult to produce close binaries with such extreme mass ratios such as those between a host star and a planet  \citep{2009MNRAS.392..590B,2011MNRAS.418..703B}. If the PMCs are formed from the gravitational collapse of the surrounding molecular cloud itself, and not formed in a circumstellar disk, we could expect the PMC to follow the same trend in planet to disk mass ratio as the parent star forming region. However, the relative disk to star/planet mass ratios do not appear to be correlated in binary systems, where the viscosity of the disk dictates the evolutionary timescales \citep[e.g.][]{2017ApJ...836..223W,2014ApJ...784...62A}.
    This mechanism is not clearly supported by the DH Tau b and GQ Lup b observations. While the circumstellar disks are detected, with median disk to stellar mass ratios indicative of a young age, the PMC disks are less massive than expected. Turbulent fragmentation is also not a good fit for FW Tau C whose disk mass is well above the disk mass for the host binary, though photoevaporation from the binary may have removed the circumbinary disk, explaining the discrepancy in the disk masses. 

\end{itemize}

Accretion rates provide another valuable indicator of formation mechanism.
 \citet{2011ApJ...743..148B} provides a picture of accretion rates for PMCs in agreement with the accretion rate-mass relation found for low-mass stars and brown dwarfs. In fact, the reported accretion rate for GSC 0614-210 B was above average when compared to a sample of similarly low mass brown dwarfs from \citet{2009ApJ...696.1589H}. Assuming these high mass accretion rates are indicative of large disk masses would seem to support formation via turbulent fragmentation. 
 In addition, it seems that most PMCs located in young ($<$ 10 Myrs), nearby star forming regions are accreting as has been seen for field brown dwarfs \citep[e.g.][]{2015A&A...579A..66M}.
 Evidence of circumplanetary disks from accretion signatures alone rejects core accretion and subsequent scattering as a possible formation pathway, as it would cause a disk to dissipate. 
If indeed future observations using a larger sample size of PMCs show that they have "normal" accretion rates for their mass but small disk masses, this could serve as a valuable marker for formation scenario.


Unfortunately, no single planet formation model is capable of explaining the observed disk masses (and upper limits) for the ensemble of known wide separation planetary mass companions. 
Nonetheless, these are very exciting results as we are likely witnessing the very first stages of gaseous planet assembly.
PMCs in general have the potential to offer unique insight into the early stages of extrasolar planet formation and to unveil, for the first time, the properties of circumplanetary disks. The observations of this type completed to date support discrepant formation scenarios. Millimeter continuum observations for more of these systems are required to pin down the mechanism capable of generating these massive companions at wide separations.

\acknowledgements
\noindent \textbf{Acknowledgements:} This material is based upon work supported by the National Science Foundation Graduate Research Fellowship under Grant No. DGE-1232825. We also acknowledge funding from ANR of France under contract number ANR-16-CE31-0013.
C.C. acknowledges support from CONICYT PAI/Concurso nacional de insercion en la academia, convocatoria 2015, Folio 79150049. M.R.S. is thankful for support from the Milenium Science Initiative, Chilean Ministry of Economy, Nucleus RC 130007 and Fondecyt (1141269). H.C. acknowledges support from the Spanish Ministerio de Econom\'ia y Competitividad under grant AYA 2014-55840-P. The authors wish to thank Karl Schuster, director of IRAM, for prompt allocation of Director observing time on the NOEMA array. MRS acknowledges support from Fondecyt (1141269) and the Millennium Nucleus RC130007 (Chilean Ministry of Economy). S.G.W. and F.M. thank Marshall Perrin for support and guidance in obtaining funding resources.


\end{document}